\documentclass[preprint]{aastex}

\usepackage{graphicx} 
\usepackage{txfonts} 
\usepackage{natbib}
\usepackage{epsfig} 
\usepackage{rotating} 
\usepackage{lscape}
\usepackage{times}

\shorttitle{Detection of H RRL Voigt Profiles toward Sgr B2(N)} \shortauthors{von Proch\'azka et al.}

\begin{document}

\title{Detection of Voigt Spectral Line Profiles of Hydrogen Radio
Recombination Lines toward Sagittarius B2(N)}

\author{Azrael A. von Proch\'azka\altaffilmark{1,2}, 
Anthony J.Remijan\altaffilmark{3,4}, 
Dana S. Balser\altaffilmark{3}, 
Robert S. I. Ryans\altaffilmark{1}, 
Adele H. Marshall\altaffilmark{5}
Fredric R.Schwab\altaffilmark{3}, 
Jan M. Hollis\altaffilmark{6}, 
Philip R. Jewell\altaffilmark{3,4}, and Frank J. Lovas\altaffilmark{4,7}}

\altaffiltext{1}{Astrophysics Research Centre, School of Mathematics and Physics, Queen's University, Belfast, BT7 1NN, UK} 
\altaffiltext{2}{Undergraduate Student, National Radio Astronomy Observatory, 520 Edgemont Road, Charlottesville, VA, 22903-2475} 
\altaffiltext{3}{National Radio Astronomy Observatory, 520 Edgemont Road, Charlottesville, VA, 22903-2475} 
\altaffiltext{4}{Center for Chemistry of the Universe, Department of Chemistry, University of Virginia, McCormick Rd., P.O. Box 4000319, Charlottesville, VA 22904-4319} 
\altaffiltext{5}{Centre for Statistical Science and Operational Research, School of Mathematics and Physics, Queen's University, Belfast, BT7 1NN, UK} \altaffiltext{6}{NASA Goddard Space Flight Center, Computational and Information Sciences and Technology Office, Greenbelt, MD, 20771} 
\altaffiltext{7}{NIST, Optical Technology Division, 100 Bureau Drive, Gaithersburg, MD, 20899}

\begin{abstract} 
We report the detection of Voigt spectral line profiles of radio recombination lines (RRLs) toward Sagittarius B2(N) with the 100-m Green Bank Telescope (GBT).  At radio wavelengths, astronomical spectra are highly populated with RRLs, which serve as ideal probes of the physical conditions in molecular cloud complexes.  An analysis of the H$n$$\alpha$ lines presented herein shows that RRLs of higher principal quantum number ($n>90$) are generally divergent from their expected Gaussian profiles and, moreover, are well described by their respective Voigt profiles.  This is in agreement with the theory that spectral lines experience pressure broadening as a result of electron collisions at lower radio frequencies.  Given the inherent technical difficulties regarding the detection and profiling of true RRL wing spans and shapes, it is crucial that the observing instrumentation produce flat baselines as well as high sensitivity, high resolution data.  The GBT has demonstrated its capabilities regarding all of these aspects, and we believe that future observations of RRL emission via the GBT will be crucial towards advancing our knowledge of the larger-scale extended structures of ionized gas in the interstellar medium (ISM). 
\end{abstract}



\section{Introduction}

Sagittarius B2(N) is a region undergoing active star formation that hosts a number of environments, including expanding H$_2$O masers (Reid et al.\ 1988); regions of large complex molecules (Snyder et al.\ 1995); in-falling, outflowing, and rotating gas (Qin et al.\ 2008); and ultra-compact and extended H{\small II} regions (Chaisson 1973; Mehringer et al.\ 1993; Miao et al.\ 1995).  Of particular interest are the radio recombination lines (RRLs) associated with these H{\small II} regions which, due to their optically thin radiation throughout the Galactic disk, may serve as accurate diagnostic probes of ionized gas at great distances.  For example, if a source exists in local thermodynamic equilibrium (LTE) and the impacts from electrons are negligible, then the line-to-continuum intensity ratio of the RRL may be used to determine the electron temperature (Shaver 1980).  Since conservation of momentum dictates that electron motions dominate the dynamics of an ionized gas, the electron temperature is considered to be a close approximation to the H{\small II} region's kinetic temperature, which may further be used to constrain the boundaries of other physical properties such as source size, flux density, rms electron density, and the magnitude of turbulence within the gas (Brown et al. 1978; Roelfsema \& Goss 1992; Rohlfs \& Wilson 1996; Gordon \& Sorochenko 2002).  Through a gradient in electron temperature over Galactocentric radius, RRLs have been used to determine the trend in Galactic metallicity (e.g., Churchwell \& Walmsley 1975; Wink et al. 1983; Shaver et al. 1983; Quireza et al. 2006).

The shape of the spectral line can also provide important information about the dynamics of the ionizing gas which constitutes the H{\small II} region.  The electrons of an H{\small II} region in LTE are expected to experience a Maxwell-Boltzmann velocity distribution and subsequently produce Gaussian line profiles via Doppler broadening.  Indeed, most published RRL profiles are well-described by Gaussian profiles.  However, for RRLs with large principal quantum numbers ($n$) arising from high density gas, pressure broadening from electron impacts is expected to be important (Griem 1967).  The profile generated by pressure broadening has a Lorentzian curve, which has higher intensity in the line wings than a Gaussian profile does.  In many cases, RRLs reflect both thermal (Doppler) and pressure broadening in the ionized gas.  Consequently, the observed shape of such RRLs is a convolution of the individual Gaussian and Lorentzian components$-$a Voigt profile.  Unfortunately, the detection of Voigt profiles from RRLs is often challenging.  The Lorentzian component is expected to become important at higher $n$, and thus lower frequencies.  Because the RRL intensity in this regime is comparatively weaker than it is at higher frequencies, the Lorentzian-broadened wings are often difficult to distinguish from instrumental baseline structure (e.g., Smirnov et al.\ 1984).  Furthermore, it is also easy to distort the RRL line shape using data reduction techniques (e.g., Bell et al.\ 2000).

While many authors have reported detection of Voigt profiles from RRL emission toward H{\small II} regions (e.g., Simpson 1973; Smirnov et al.\ 1984; Foster et al.\ 2007), few have obtained the high spectral resolution, high sensitivity observational data necessary to achieve unequivocal results.  The difficulty in obtaining such data is due to several factors: (i) low signal-to-noise ratios (S/N) for which the enhanced line wings of the Voigt profile cannot be detected; (ii) poor spectral resolution, which prevents sufficient sampling of the RRL profile; (iii) the presence of nearby RRLs (e.g., He$\alpha$, as seen in Figs. 1$-$4), which produce confusion within the spectrum; (iv) the broad beams of many telescopes (especially single-dish ones), which dilute and mix regions dominated by Doppler broadening with those which contain significant pressure broadened components; (v) large-scale motions, such as infalling and outflowing ionized gas, which produce asymmetric wings to appear on a thermal (Gaussian) profile; and (vi) poor spectral baselines.  The last factor has arguably been the most difficult to circumvent as instrumental effects can distort the spectral baselines, making it very difficult to distinguish the line wings from the baseline (e.g., Bania et al.\ 1997).  Nevertheless, there is ample evidence that pressure broadening is important in H{\small II} regions.  In particular, non-LTE radiative transfer models can only match the observed RRL intensity or line width for some H{\small II} regions if pressure broadening is included (e.g., Churchwell 1971; Balser et al.\ 1999; Keto et al.\ 2008).

To our knowledge, the only unambiguous detection of RRL Voigt profiles are at low radio frequencies ($<$100 MHz) where the lines are detected in absorption toward a bright continuum object. Kantharia et al.\ (1998) detected Voigt profiles of carbon RRLs toward Cas A with an effective integration time of 400 hours.  More recently, Stepkin et al.\ (2007) detected carbon RRLs toward Cas A from the largest bound atoms in space with $n\approx1009$ and an integration time of 504 hours.  These studies were able to detect Voigt profiles for several reasons: (i) the very long integration times provided the necessary sensitivity; (ii) the Lorentzian component of the Voigt profile is more pronounced at the lower frequencies observed; and (iii) the bright background intensity from Cas A and the Galactic background produces radiation broadening, which also contributes to the Lorentzian component.  Measuring the Voigt profile was critical in constraining the physical conditions in these regions, which are thought to be associated with the cold atomic hydrogen component of the diffuse ISM.

The design of the GBT, which leaves its aperture unblocked, provides many improvements in sensitivity, spectral resolution, and spectral baselines over traditional single-dish telescopes.  For example, multi-path reflections from the superstructure are avoided by the GBT (Fisher et al.\ 2003).  Although much of the instrumental baseline structure produced by radio telescopes is correlated out in radio interferometers, until recently most instruments did not have sufficient spectral resolution to resolve Voigt profiles. 

In this article we report on the clear detection of (at least) four emission line RRL Voigt profiles from observations directed toward Sagittarius B2(N) as part of the Prebiotic Interstellar Molecule Survey (PRIMOS\footnote{For more information on obtaining all the raw and processed data and a full summary of the observational parameters, visit http://www.cv.nrao.edu/$\sim$aremijan/PRIMOS}).  A sample of the 12 highest S/N RRLs detected in this survey are presented, with frequencies ranging from 4.7 to 24.5 GHz, corresponding to principal quantum number $n$ between 64 and 111.  All told, there will be about 1200 H RRLs present within the PRIMOS survey frequency range of 1-50 GHz.  This includes only those RRLs from the $\Delta$$n$=1 (alpha lines) to $\Delta$$n$=6 (zeta lines) series.  The RRL rest frequencies were taken from the Splatalogue\footnote{Available at http://www.splatalogue.net.} database (Remijan et al.\ 2007).  It is beyond the scope of this paper to discuss all the currently detected RRLs toward the Sagittarius B2(N) region.  Rather, the main point of this work is to present the data repository where researchers can obtain the full complement of RRLs detected and to illustrate that Voigt spectral-line profiles of RRLs can clearly be detected with the GBT.  As such, we discuss only a subset of the data available and illustrate that high-$n$ ($n>$90) H RRLs toward Sagittarius B2(N) are best fit with Voigt spectral line profiles.

Figures 1$-$4 and Tables 1$-$2 summarize our results.  Spectra of 12 H$n$$\alpha$ RRLs are shown with Gaussian, Lorentzian, and Voigt function profile fits to the raw data.  An expanded view of the line wings near the baseline is shown to illustrate the differences in these models.  All spectra are plotted with respect to the local standard of rest (LSR) velocity, centered at 0 km s$^{-1}$.  Table 1 lists the fitting parameters for each of the H$n$$\alpha$ RRLs.  We present a review of the observations made in \S{2}, describe our reduction methods and analysis in \S{3}, and discuss the data and profile fits in \S{4}.

\section{Observations}

All observations were obtained between 4.7 and 24.5 GHz with the NRAO\footnote{The National Radio Astronomy Observatory is a facility of the National Science Foundation, operated under cooperative agreement by Associated Universities, Inc.} 100-m Robert C. Byrd Green Bank Telescope from 2004 March 4 to 2005 November 12.  Briefly, the data were obtained using the GBT spectrometer, configured to provide four intermediate frequency spectral windows simultaneously in two orthogonal polarizations through the use of offset oscillators.  The spectral resolution was 24.4 kHz at frequencies centered between 4.7--24.5 GHz.  This is significantly better than many previous measurements\footnote{For example, Smirnov et al (1984) had a spectral resolution of 63, 115, and 550 kHz centered at observing frequencies of 4.8-9.1 GHz, 22.4 GHz, and 36.5 GHz, respectively using the 100-m Effelsberg telescope.}.

The GBT's half-power beam-width can be approximated by: $\theta_b \approx \frac{740''}{\nu}$, where the observing frequency ($\nu$) is measured in GHz.  Over the course of the observations, the spectrometer bandpasses were centered at different frequencies in the range of 4.7 GHz to 24.5 GHz that correspond to $\theta_b \approx 157''$ and $\theta_b \approx 30.2''$, and a beam efficiency from 0.96 to 0.83, respectively.  The aperture efficiency is 0.70 to 0.61 over the same frequency range\footnote{Bell et al. (2000) estimated $\eta_B = 0.4 \pm\ 0.04$ and $\eta_A = 0.3 \pm\ 0.03$ at a frequencies of 6.0 and 17.6 GHz, respectively using the NRAO 140-ft telescope.}.

The J2000 pointing position employed was $\alpha$ = 17$^h$47$^m$19$^s$.8, $\delta$ = -28$^o$22$'$17$"$ ($\ell=0.677^\circ, b=-0.027^\circ$), which is centered on the Sagittarius B2(N-LMH) source and an LSR source velocity of +64 km s$^{-1}$ was assumed.  The dynamic pointing and focusing corrections were automatically updated and employed based on realtime temperature measurements of the structure input to a thermal model of the GBT; local pointing and focus corrections were adjusted typically every two hours using standard pointing sources.  Spectrometer data were taken in the OFF-ON position-switching mode, with the OFF position at 60$'$ East in azimuth with respect to the ON source position.  A single scan consisted of two minutes in the OFF source position followed by two minutes in the ON source position.  Sagittarius B2(N) was observed in this manner above 10$^{\circ}$ elevation from source rise to source set (a six-hour track).  When available, all the accumulated scans of the two polarization outputs from the spectrometer were averaged in the final data-reduction process to improve the S/N.  The antenna temperatures of the RRLs presented in this paper are on the T$_A^*$ scale (Ulich \& Haas 1976) with estimated 20\% uncertainties.

\section{Reduction and Analysis} 

Before any profile fitting was done, the continuum level was subtracted from each spectrum, using up to a third-order polynomial fit to the baseline.  This removed any instrumental slopes in the bandpass so that the baselines would be flat enough for a profile analysis, yet the shape of the original RRL would be preserved.  All RRLs were processed with the NRAO GBTIDL data reduction package according to standard reduction protocol (see, e.g., Garwood et al. 2007).  Particular care was taken to preserve the shape of the line wings.

The equations used to generate Gaussian and Lorentzian profile fits for the RRLs in this survey were 

\begin{equation}
\phi_G(\nu)=A_{0} \exp{\left(-{\frac{(\nu-\nu_{0})^2}{2 \sigma^2}}\right)},
\end{equation}

\noindent
where the Gaussian FWHM is given as 

\begin{equation}
{\rm FWHM_G}(\sigma)=2 \sigma \sqrt{2 \ln(2)}
\end{equation}

\noindent
and 

\begin{equation}
\phi_L(\nu)= A_{0} Ê\frac{\gamma^2}{{(\nu-{\nu_0})^2} + {{\gamma}^2}},
\end{equation}

\noindent
where the Lorentzian FWHM is given as 

\begin{equation}
{\rm FWHM_L}(\gamma)=2 \gamma, 
\end{equation}

\noindent where A$_0$ represents line intensity, $\nu_0$ represents frequency at line center, $\nu$ is the varying frequency across the breadth of the RRL, $\sigma$ is the one-half value at 1/$e$ intensity of the Gaussian line profile, and $\gamma$ is one-half value at 1/2 intensity of the Lorentzian line profile.

In general, astronomical environments produce RRLs that experience both Doppler and pressure broadening.  However, depending on the RRL transition and the physical properties of the gas, one type of broadening mechanism is usually observed to dominate over the other.  At radio frequencies, Doppler broadening is almost always measured.  In cases for which it is possible to detect both Doppler and pressure broadening, the observed spectral line profile is expected to be a convolution between the Gaussian and Lorentzian profiles.  The convolved shape is referred to as a Voigt profile and may be described by the formula 

\begin{equation} \phi_V(\nu)=\frac{A_{0} e^{{-\frac{\gamma^2}{2
\sigma^2}}} \left(e^{{-\frac{(\nu-\nu_{0}+i \gamma)^2}{2 \sigma^2}}}
{\rm Erfc} \left[-{\frac{i(\nu-\nu_{0})+ \gamma}{\sqrt{2}
\sigma}}\right] + e^{{-\frac{(\nu-\nu_{0}-i \gamma)^2}{2 \sigma^2}}}
{\rm Erfc} \left[{\frac{i(\nu-\nu_{0})+ \gamma}{\sqrt{2}
\sigma}}\right]\right)}{2 {\rm Erfc} \left[-{\frac{\gamma}{\sqrt{2}
\sigma}} \right]}, \end{equation} 

\noindent where Erfc is the complementary error function. 

Each H RRL presented in this article was fitted with Gaussian, Lorentzian, and Voigt profiles produced by the $Mathematica$ {\tt NonLinearRegress} least-squares-fitting routine.  The various fit parameters and error estimates for each RRL are summarized in Table 1.  Gaussian, Lorentzian, and Voigt models were generated using Equations (1), (3), and (5), respectively.  For the Gaussian and Lorentzian fits we list the peak intensity, the FWHM line width, and the sum-of-squares (SOS) residual value, which measures the deviation of each point from the best fit model.  For the Voigt fit we list the peak intensity, the Gaussian and Lorentzian FWHM line widths, and the SOS residual value.  Also listed is the ratio $a = FWMH_L/FWMH_G$, which is the ratio of the Lorentzian-to-Gaussian component FWHM values and illustrates the degree of deviation of the line profile from a pure Gaussian or pure Lorentzian line shape.  The estimates of the model parameters were found from a least-squares fit, minimizing the SOS residuals, and the model parameters are reported over a 99.7\% confidence interval, representative of a 3$\sigma$ error on the fit.  These errors, reported in Table 1, are on the last significant figure.  In every case, the Voigt fit gives the smallest SOS residual, suggesting that it may be the best fit profile for the observed data.

Figure 1 contrasts an apparently Gaussian profile (the H69$\alpha$ transition at 19590 MHz) with a strongly Voigt-shaped profile (the H109$\alpha$ transition at 5009 MHz).  For the H69$\alpha$ line, the best fit Voigt profile is almost exactly coincident with the Gaussian profile, indicating that pressure broadening is not significant for this case.  In contrast, for the H109$\alpha$ profile, the line intensity and shape clearly require contributions from both pressure and Doppler broadening to match the data.  Figures 2$-$4 show the RRL baseline wings with the various profile fits in order of decreasing $n$.  The evolution in line shape is clear, showing the shift from predominantly Voigt-shaped profiles at lower frequencies (4744, 5009, 5444, and 8584 MHz) to more Gaussian-shaped profiles at higher frequencies.  Physically, this corresponds to a transition from lines broadened by a combination of both Doppler motions and electron impacts to lines broadened primarily by Doppler motions alone.  The wings of the highest frequency RRL transition (24510 MHz) appear from visual inspection to be well-modeled by a Gaussian profile.  However, between the 8873 MHz and 19590 MHz transitions, the RRL profiles appear to fluctuate between either having shapes mostly described by Gaussian profiles (14129, 15281 MHz) and having wing shapes and breadths intermediate to those expected for Gaussian and Voigt profiles (9173, 10161, 17992, and 19590 MHz).  As a general trend, we see support of the theory that for higher values of $n$, pressure broadening becomes increasingly important.  

Further statistical analysis was carried out using custom code within the SAS environment to compare the observed data to the Voigt, Gaussian, and Lorentzian fits for each of the H$n$$\alpha$ RRLs listed in Table 1.  Correlation analyses between the observed data and that from the three distributions at all frequencies indicate in every case a strong correlation between the observed data and the data generated from the fitted statistical distribution (p $<$ 0.0001).  The median test (Table 2) examines whether there is a significant difference between the observed data and the generated data by counting the number of values in the data set above and below the calculated median and statistically comparing the frequencies.  If the p-value is less than 0.05, this implies that there is a statistically significant difference between the observed data and the data from the proposed distribution.  If the p-value is greater than 0.05, on the other hand, this implies that the two data sets come from the same distribution.  For all H RRL frequencies, the Voigt distribution shows that there is no significant difference between the observed data and the Voigt fit.  This is particularly the case with the lowest frequencies (4744, 5009, 5444, 8584, and 8873 MHz), where the test suggests that there is no evidence at all that the observed data and the Voigt data are not from the same distribution.  In fact, the Voigt data are considered the best representation for the observed data at all frequencies except 10161 MHz, where the Gaussian data are considered better.  We note that the 10161 MHz line appears to suffer from considerable spectral contamination on its positive velocity wing and that in all cases, the presence of the corresponding He$\alpha$ RRL interferes with the fit on the negative velocity side of the H$\alpha$ RRL.  The analysis also indicates that the Lorentzian distribution is a significantly poor fit (p $<$ 0.05) for all frequencies except 4744 MHz, where it is not as good as the Voigt distribution (p$=$0.3323 compared to p$=$0.99999) but that it is statistically better than the Gaussian distribution (p$=$0.0878). 

The median test, however, is less robust than the Wilcoxon Mann Whitney U test (Table 2), which not only considers the values that differ from the median but their also their magnitudes.  Examination of the Wilcoxon statistics indicates that again the Voigt is the better choice for the lower frequencies (4744, 5009, 5444, 8584 MHz) but not necessarily so for the higher frequencies.  For the profiles at 8873 MHz and above, the Voigt data are significantly different from the observed data.  The Gaussian data are considered statistically closer to the observed data for the 14129, 15281, and 24510 MHz profiles.  However, it is apparent that the remaining frequencies (9173, 10161, 17922, and 19590 MHz) are not suitably represented by any of the three distributions as the p-value is always less than 0.05, suggesting that the observed data are significantly different from the distributions data for these frequencies.  The analyses also confirm previous results that the Lorentzian distribution is a poor fit for all frequencies except 4744 MHz, where it is not as good as the Voigt (p$=$0.1211 compared to p$=$0.4924) but is statistically better than the Gaussian (p$=$0.0006).

\section{Discussion}

RRL theory predicts that emission from ionized gas will produce Voigt line profiles, which consist of a Doppler (Gaussian) component and a pressure broadened (Lorentzian) component (Griem 1967).  This assumes that no large-scale motions such as expansion or rotation exist within the observed region.  Pressure broadening is caused by impacts from electrons and depends on the local electron density and the principle quantum number $n$.  Pressure broadening is a sensitive function of $n$; as $n$ increases, the line profile is expected to go from being dominated primarily by the Gaussian component to being affected by a significant Lorentzian component as well (Brocklehurst \& Seaton 1972).  This trend can be detected in several ways. Since the Lorentzian profile decays as a power law, rather than exponentially as the Gaussian does, RRL profiles in regions of ionized gas for which pressure from electron collisions is significant are expected to have increasingly broadened wings toward their baselines (see Figures 2$-$4).  As theorized by Griem (1967), the amount of pressure broadening present can be quantitatively measured by the ratio of Lorentzian-to-Gaussian influences on the spectral line's FWHM value (i.e. parameter $a$ in Table 1).  By inspecting these values, we see that it does indeed generally increase with $n$.  That is to say, the Lorentzian component becomes significantly dominant, which allows the wings of the Voigt profile to be detected.  That these particular RRLs are indeed Voigt in nature is confirmed through both median and Wilcoxon statistical tests.

Since the frequency of an RRL is a function of $n$, the telescope beam size (and possibly the opacity as well) change with $n$.  Detailed models of the electron temperature and density are required for a proper physical analysis of RRLs such as those reported here.  Such a rigorous test of pressure broadening on a source as complex as Sagittarius B2(N) is beyond the scope of this paper and might prove challenging with the current data set.  Nevertheless, the expected theoretical trends of pressure broadening appear to be correct, and we conclude from visual and quantitative analyses that at least four of the RRLs detected in this survey$-$4744, 5009, 5444, and 8584 MHz$-$are verifiably Voigt in nature.  With today's spectrometers, sacrificing velocity resolution for more channels (leading to poor velocity resolution) is no longer an issue against detecting extended wings in line profiles.  Spectrometer improvements to existing (e.g., EVLA) and future (e.g., ASKAP) interferometers hold much promise as probes of pressure broadening within high density, compact H{\small II} regions, and data achievable with the GBT will certainly provide a crucial role towards advancing our understanding of the larger-scale, extended ionized structures in the ISM.


\section{Acknowledgements}

We would like to thank M. J. Remijan for continuing programming support and development and an anonymous referee for valuable comments. This work supported in part by the NSF Centers for Chemical Innovation through award CHE-0847919.

\begin{sidewaystable}
\centering 
\resizebox{8.0in}{!} { 
\begin{tabular}{c| c| c c c c c | c c c  | c c c} 
\hline 
\hline 
&   & \multicolumn{5}{c}{Voigt} & \multicolumn{3}{c}{Gaussian}  & \multicolumn{3}{c}{Lorentzian} \\
\hline
\hline
$\nu$   &n      &A$_V$   &FWHM(G) &FWHM(L)   &$a$  &SOS(V)  & A$_G$  &FWHM$_G$ &SOS(G)   &A$_L$    & FWHM$_L$     &SOS(L)\\ 
\hline
4744    &111 &0.98(1)   &11.2(6)      &30.1(4)        &2.69 & 0.060    &0.92(2)   &24.7(4)          &0.340       &1.02(1)   & 38.8(3)             &0.103\\ 
5009    &109 &0.95(5)   &12.9(19)    &29.3(25)     &2.27 & 1.174    &0.89(4)   &28.0(5)          &1.445       &0.98(5)   & 37.9(9)             &1.215\\ 
5444    &106 &1.16(2)   &13.7(5)      &27.9(8)        &2.04 & 0.261    &1.09(2)   &27.9(2)          &0.671       &1.20(2)   & 37.7(3)             &0.350\\ 
8584    &91   &1.73(2)   &17.0(3)      &15.8(5)        &0.93 & 0.531    &1.68(2)   &24.1(1)          &1.052       &1.84(3)   & 32.5(3)             &1.453\\ 
8873    &90   &1.51(5)   &16.8(8)      &15.3(15)      &0.91 & 4.215    &1.46(4)   &23.6(3)          &4.595       &1.60(5)   & 32.0(5)             &4.935\\ 
9173    &89   &1.70(2)   &16.8(3)      &15.4(6)        &0.92 & 0.811    &1.65(2)   &23.7(1)          &1.329       &1.81(3)   & 32.0(5)             &1.763\\ 
10161  &86   &1.69(6)   &14.2(9)      &18.8(16)     &1.32 & 4.448    &1.63(6)   &22.7(3)          &5.334       &1.77(6)   & 31.5(6)             &5.099\\ 
14129  &77   &1.68(1)   &16.0(3)      &7.0(6)          &0.44 & 0.817    &1.66(1)   &18.5(1)          &0.983       &1.81(3)   & 29.1(3)             &3.308\\ 
15281  &75   &1.42(1)   &16.1(4)      &6.2(6)          &0.39 & 0.790    &1.40(1)   &18.2(1)          &0.892       &1.53(2)   & 28.8(3)             &2.840\\ 
17992  &71   &1.83(2)   &15.7(3)      &7.3(5)          &0.46 & 1.660    &1.80(1)   &18.3(1)          &1.947       &1.97(3)   & 28.9(3)             &5.360\\ 
19590  &69   &1.61(1)   &15.6(4)      &7.2(5)          &0.46 & 1.223    &1.59(1)   &18.2(1)          &1.456       &1.74(3)   & 28.8(3)             &4.369\\ 
24510  &64   &1.31(1)   &15.6(2)      &6.1(6)          &0.39 & 1.461    &1.29(1)   &17.7(1)          &1.600       &1.41(3)   & 28.1(3)             &4.112\\ 
\hline 
\hline 
\end{tabular} }
\caption{\footnotesize{Parameters of the best fit Gaussian, Lorentzian, and Voigt profiles of the selected RRLs. Note $-$ $\nu$ represents the approximate RRL frequency in MHz, and $n$ represents the principal quantum number of the RRL observed.  A$_V$,  A$_G$, and A$_L$ represent the line intensities of the Voigt, Gaussian, and Lorentzian profiles, FWHM(G) and FWHM(L) represent the respective contributions of the Gaussian and Lorentzian line widths to the Voigt line width. $a$ is the ratio of line widths: FWHM(L)/FWHM(G). SOS(V), SOS(G), and SOS(L) are the sum-of-squares values for the Voigt, Gaussian, and Lorentzian profiles.  FWHM$_G$ and FWHM$_L$ represent line intensity and FWHM values for the Gaussian and Lorentzian profiles.  Values given in parentheses are the 3$\sigma$ confidence intervals of the fits.}}
\end{sidewaystable}

\begin{table}
\centering 
\resizebox{6.0in}{!} { 
\begin{tabular}{c| c| c| c| c| c| c} 
\hline 
\hline 
Frequency  & \multicolumn{3}{c}{Median test} & \multicolumn{3}{c}{Wilcoxon test}\\
\hline
\hline
           & Gaussian & Lorentzian & Voigt & Gaussian & Lorentzian & Voigt \\ 
\hline
4744   & 0.0878    & 0.3293    & 1.0000 & 0.0006    & 0.1211    & 0.4924    \\
5009   & 0.0028    & 0.0028    & 0.4388 & $<$0.0001 & 0.0271    & 0.3346    \\
5444   & $<$0.0001 & 0.0242    & 0.4124 & $<$0.0001 & 0.0146    & 0.3884    \\
8584   & $<$0.0001 & $<$0.0001 & 0.3397 & $<$0.0001 & $<$0.0001 & 0.2444    \\
8873   & 0.0475    & 0.0003    & 0.5889 & $<$0.0001 & $<$0.0001 & 0.0121    \\
9173   & 0.0001    & $<$0.0001 & 0.3984 & 0.0048    & $<$0.0001 & 0.0193    \\
10161  & 0.5549    & 0.0304    & 0.4311 & 0.0419    & $<$0.0001 & 0.0041    \\
14129  & 0.1959    & $<$0.0001 & 0.5943 & 0.1192    & $<$0.0001 & 0.0005    \\
15281  & 0.0777    & $<$0.0001 & 0.1799 & 0.1799    & $<$0.0001 & $<$0.0001 \\
17922  & 0.0832    & $<$0.0001 & 0.4240 & $<$0.0001 & $<$0.0001 & 0.0023    \\
19590  & 0.0832    & $<$0.0001 & 0.4463 & $<$0.0001 & $<$0.0001 & 0.0004    \\
24510  & 0.4932    & $<$0.0001 & 0.7555 & 0.3094    & $<$0.0001 & $<$0.0001 \\

\hline 
\hline 
\end{tabular} }
\caption{Fit quality statistics for various RRL model types.  Note$-$Strong correlation (p $<$  0.0001) was measured between each observed RRL and its corresponding statistical fit.  According to the median test, the Voigt function produces the best fit profile in every instance except for the RRL at 10161 MHz.  According to the Wilcoxon test, the RRLS at 4744, 5009, 5444, and 8584 MHz are best described by the Voigt profile, the RRLs at 14129, 15281 and 24510 MHz are best described by the Gaussian profile, and the RRLs at 9173, 10161, 17922, and 19590 MHz are not well-described by any of the three distributions investigated.  See \S{3} for details.}
\end{table}

\begin{sidewaysfigure} 
\begin{center} 
\epsfig{file=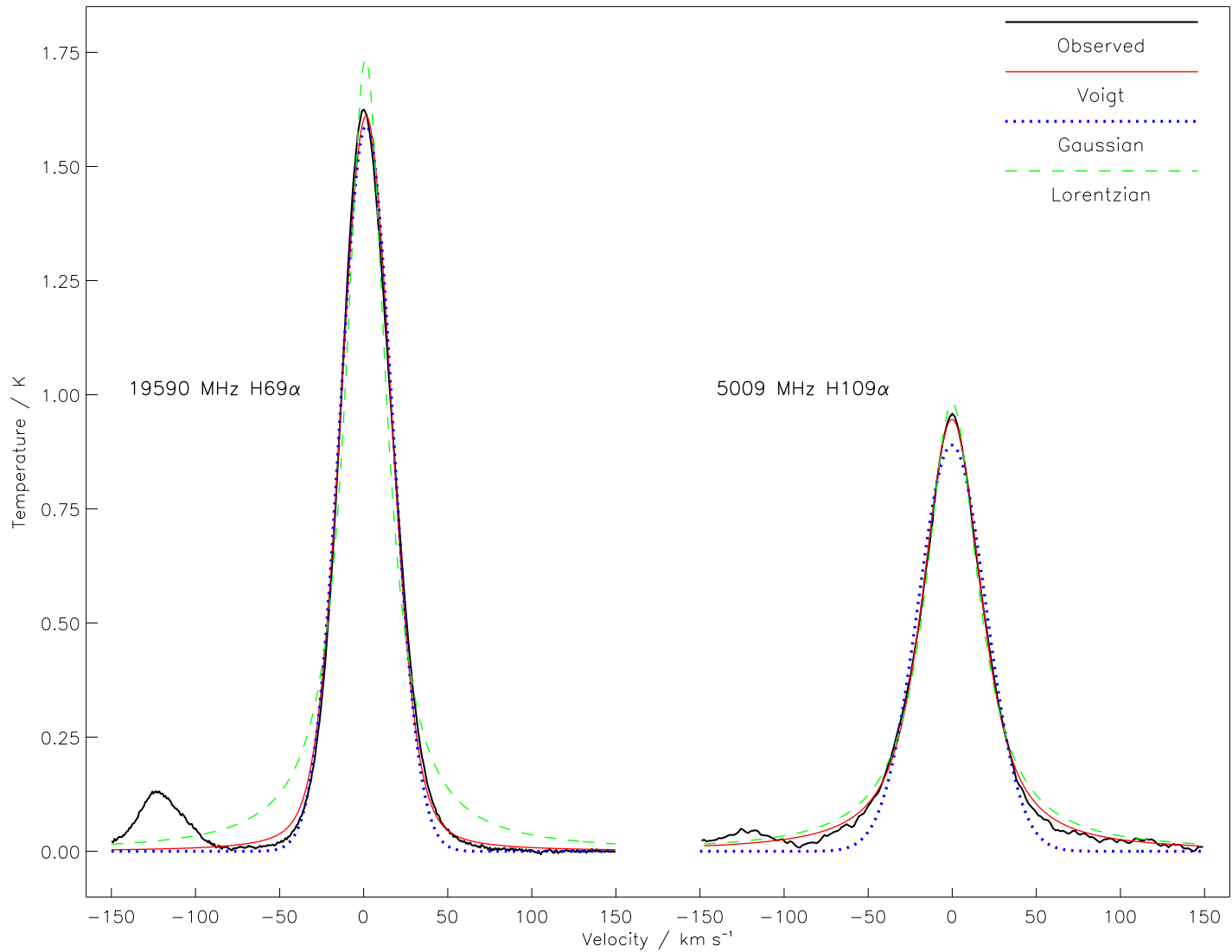, height=6.0in} 
\caption{\footnotesize{The observed H 109$\alpha$ RRL at 5009 MHz (right) and H 69$\alpha$ RRL at 19590 MHz (left) from observations taken towards Sagittarius B2(N) illustrating the shift from nearly a pure Doppler (Gaussian) line profile to one with a substantial amount of pressure broadening (Lorentzian).  In both cases, the Voigt profile has a lower SOS residual value than either the Gaussian or the Lorentzian profile.  The observed data are shown in the bold trace.  The best Voigt fit is shown as a solid line, the best Gaussian fit is shown as a dotted line, and the best Lorentzian fit is shown as a dashed line.  The weaker line to the negative velocity side of the H RRL is the corresponding He RRL.}}
\end{center} 
\end{sidewaysfigure}

\begin{figure}
\begin{center} 
\epsfig{file=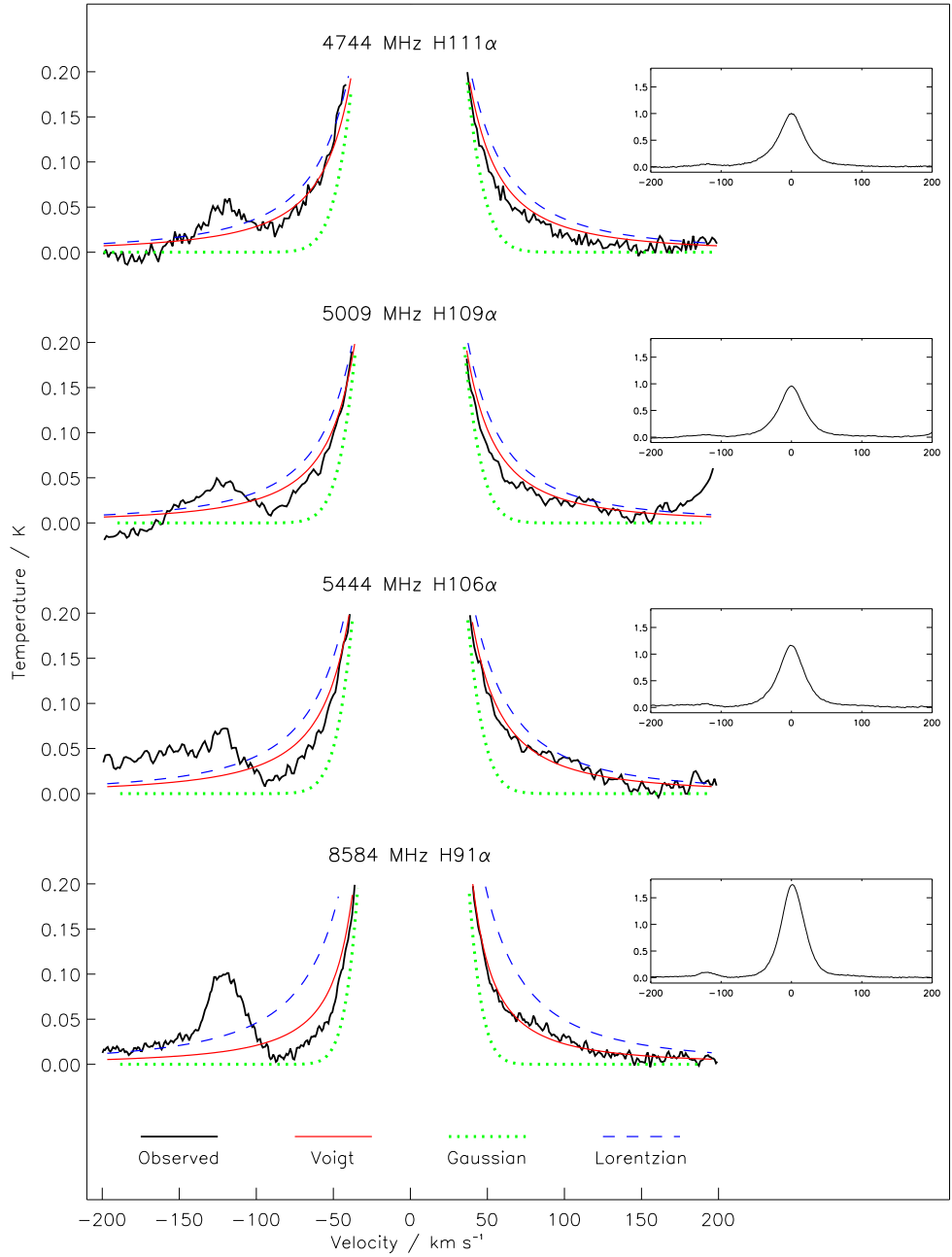, height=7.0in}
\caption{\footnotesize{Wings of RRLs between 4744 MHz and 8584 MHz.  View of full raw data is inset (y-axis ranges from 0 to 1.5 K, x-axis ranges from -200 to 200 km s$^{-1}$).  The observed data are shown in the bold trace.  The best Voigt fit is shown as a solid line, the best Gaussian fit is shown as a dotted line, and the best Lorentzian fit is shown as a dashed line.  In all four cases, the best statistical fit to the data is the Voigt profile.}}
\end{center} 
\end{figure}


\begin{figure} 
\begin{center} 
\epsfig{file=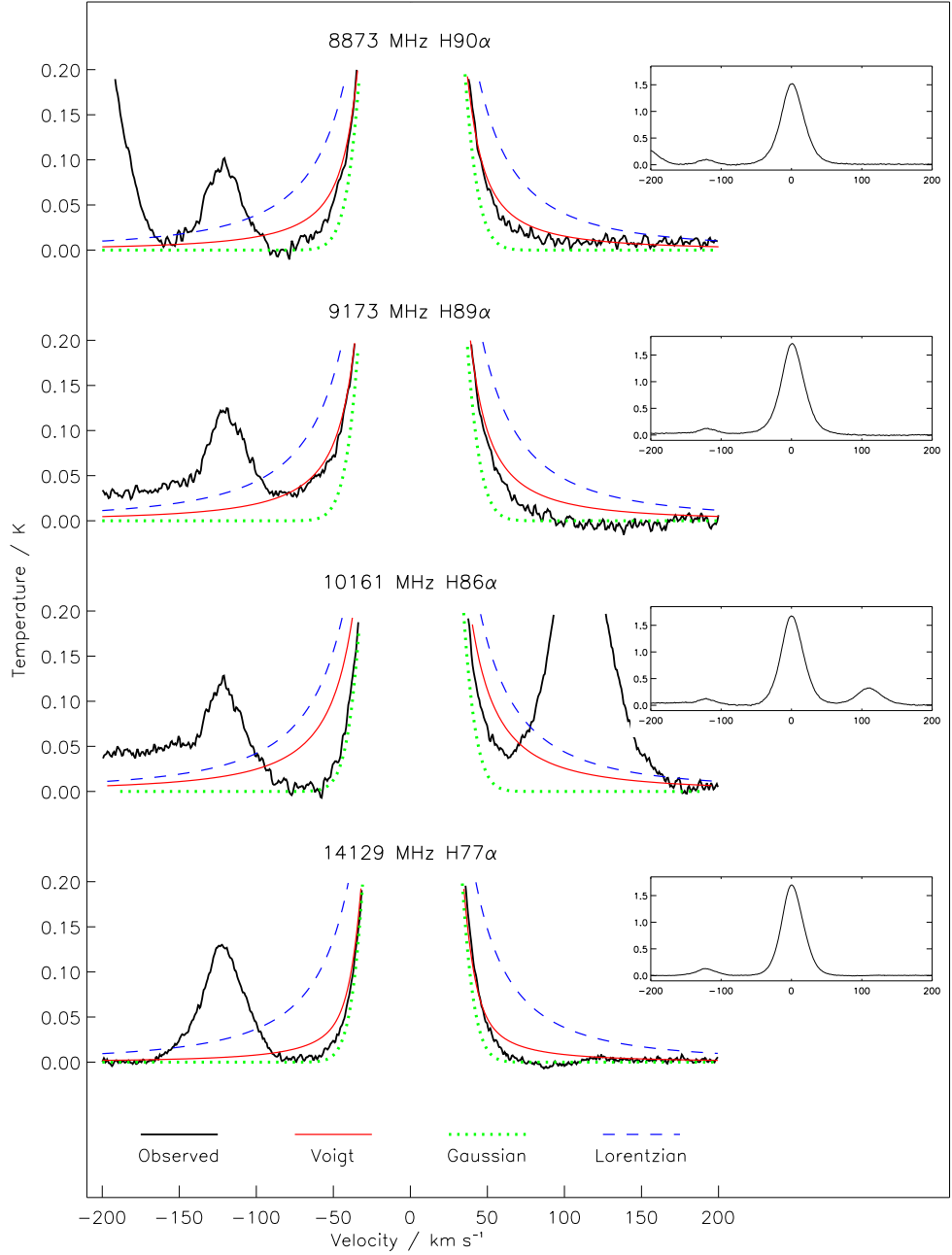, height=7.0in}
\caption{\footnotesize{Wings of RRLs between 8873 MHz and 14129 MHz.  View of full raw data is inset (y-axis ranges from 0 to 1.5 K, x-axis ranges from -200 to 200 km s$^{-1}$).  Figure labels, see Figure 2.}}
\end{center} 
\end{figure}

\begin{figure} 
\begin{center} 
\epsfig{file=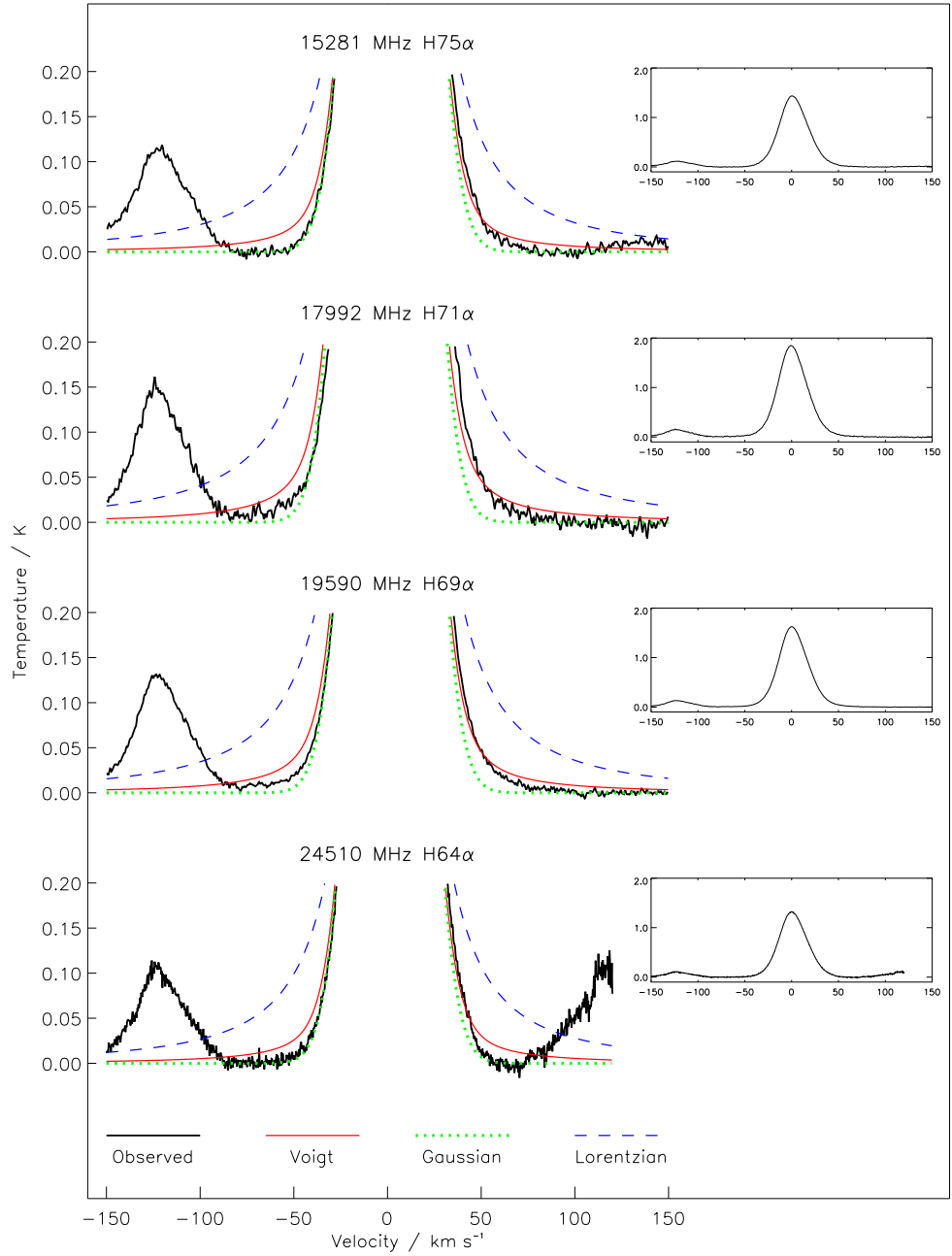, height=7.0in}
\caption{\footnotesize{Wings of RRLs between 15281 MHz and 24510 MHz. View of full raw data is inset (y-axis ranges from 0 to 2.0 K, x-axis ranges from -150 to 150 km s$^{-1}$).  Figure labels, see Figure 2.}}
\end{center} 
\end{figure}

{}

\end{document}